\documentclass{article}
\usepackage{spconf,amsmath,graphicx,hyperref}
\usepackage{cite}
\usepackage{amsmath,amssymb,amsfonts}
\usepackage{bm}
\usepackage{algorithmic}
\usepackage{graphicx}
\usepackage{float}
\usepackage{subfigure}
\usepackage{subcaption}
\usepackage[labelformat=parens]{subcaption} % 导言区加载
\usepackage{textcomp}
\usepackage{xcolor}
\usepackage{booktabs}
\usepackage{array}
\usepackage{tabularx}
\usepackage{color}
\makeatletter
\renewcommand{\maketag@@@}[1]{\hbox{\m@th\normalsize\normalfont#1}}%
\makeatother

\title{Non-Asymptotic Performance Analysis of DOA Estimation Based on Real-Valued Root-MUSIC}
\name{Junyang Liu$^{1}$, Weicheng Zhao$^{2}$, Qingping Wang$^{3,\dagger}$, Xiangtian Meng$^{4}$, Maria Greco$^{5}$, Fulvio Gini$^{5}$ \thanks{$^\dagger$Corresponding author: Qingping Wang}}
  
\address{$^1$ Department of Computing, The Hong Kong Polytechnic University, Hong Kong, China \\
      $^2$ School of Electronics and Information Engineering, Harbin Institute of Technology, Harbin, China \\
      $^3$ State Key Laboratory of CEMEE, National University of Defence Technology, Changsha, China \\
      $^4$ School of Information Science and Engineering, Harbin Institute of Technology, Weihai, China \\
      $^5$ Department of Information Engineering, University of Pisa, Pisa, Italy
      }
 
\begin{document}
%\ninept
%
\maketitle
\begin{abstract}
This paper presents a systematic theoretical performance analysis of the Real-Valued root-MUSIC (RV-root-MUSIC) algorithm under non-asymptotic conditions. 
% However, RV-root-MUSIC suffers from the problem of estimation ambiguity for the mirror roots, therefore the conventional beamforming (CBF) technique is typically employed to filter out the mirror roots. Through the equivalent subspace based on the conjugate extension method and the equivalence of perturbations for both true roots and mirror roots , this paper provides a comprehensive investigation of three critical aspects: noise subspace perturbation, true root perturbation, and mirror root perturbation characteristics in the RV-root-MUSIC algorithm. The statistical model is established and the generalized expression of perturbation is further developed.
% The simulation results show the correctness and validity of the derived statistical characteristics. The results provide a solid theoretical foundation for optimizing the parameter selection of Direction-of-Arrival (DOA) estimation in practical applications, particularly in radar systems, communication networks, and intelligent sensing technologies.
A well-known limitation of RV-root-MUSIC is the estimation ambiguity caused by mirror roots, which are typically suppressed using conventional beamforming (CBF). By leveraging the equivalent subspace constructed through the conjugate extension method and exploiting the equivalence of perturbations for true and mirror roots, this work provides a comprehensive study of three key aspects: noise subspace perturbation, true-root perturbation, and mirror-root perturbation. A statistical model is established, and generalized perturbation expressions are derived. Monte Carlo simulations confirm the correctness and effectiveness of the theoretical results. The analysis provides a rigorous foundation for parameter optimization in Direction-of-Arrival (DOA) estimation, with applications in radar, wireless communications, and intelligent sensing.
\end{abstract}
\section{Introduction}
\label{sec:intro}
Direction-of-Arrival (DOA) estimation is a fundamental nonlinear parameter estimation problem. Accurate parameter acquisition is essential for reliable target perception in diverse applications, including radar detection, wireless communications, sonar localization, and medical imaging \cite{intro1,intro2,intro3,intro4,intro5,iotj2024,iotj2025}. However, unavoidable noise contamination during signal transmission and reception introduces estimation errors, making accuracy a key criterion for evaluating algorithmic performance. Developing robust error quantification frameworks and systematic performance analysis methodologies is therefore of significant importance \cite{TAP}.

The Multiple Signal Classification (MUSIC) algorithm \cite{MUSIC} marked a breakthrough in DOA estimation by achieving super-resolution performance. However, its reliance on exhaustive spectral peak searches results in high computational complexity, which limits practical applicability. To address this issue, the root-MUSIC algorithm \cite{root-MUSIC} was introduced, significantly reducing computational burden. Nonetheless, both MUSIC and root-MUSIC operate in the complex domain, which can still pose efficiency challenges \cite{Low}. To further improve efficiency, real-valued subspace-based approaches such as Real-Valued MUSIC (RV-MUSIC) \cite{RV} and Real-Valued root-MUSIC (RV-root-MUSIC) \cite{FastDOA,RV-root-MUSIC} have been proposed. While root-MUSIC performs well under high signal-to-noise ratio (SNR) and asymptotic conditions, its accuracy deteriorates considerably in low-SNR or limited-snapshot scenarios, i.e., non-asymptotic conditions. Although performance analyses for conventional complex-domain DOA estimation algorithms exist \cite{LiFu,PA-of-root}, real-valued methods have not been thoroughly examined.

This paper addresses this gap by presenting a non-asymptotic performance analysis of RV-root-MUSIC. Theoretical error expressions are derived for both true and mirror roots, and their validity is verified through Monte Carlo simulations, which demonstrate close agreement between theoretical predictions and empirical results.

\section{PRELIMINARIES}
\label{sec:format}
\subsection{The model of the signal and arrays}
Considering $K$ far-field narrowband signals arriving on uniform linear arrays (ULA) with $L$ elements, the ideal received array signal output at $M$ snapshots can be expressed as:
\begin{equation}
\bm{X}(t)=\bm{A}(\theta)\bm{S}(t)
\end{equation}
where $\bm{X}(t)\in \mathbb{C}^{L\times M}$ is the ideal received signal matrix, $\bm{A}(\theta )\in \mathbb{C}^{L\times K}$ is the array manifold matrix of ULA, consisting of steering vectors as its columns, expressed as:
\begin{equation}
\bm{a}(\theta _k)=\begin{bmatrix}
1 & e^{-j\frac{2\pi d}{\lambda }sin\theta_k } & \cdots  & e^{-j(L-1)\frac{2\pi d}{\lambda }sin\theta_k }\\
\end{bmatrix}^T
\end{equation}
where $d$ denotes the inter-element spacing, and $\lambda$ is the signal wavelength. $\bm{S}(t)\in \mathbb{C}^{K\times M}$ is the source signal matrix.

When the perturbation $\Delta \bm{X}$ is taken into account, the above expression can be rewritten as:
\begin{equation}
\bm{\tilde{X}}(t)=\bm{X}(t)+\Delta \bm{X}=\bm{A}(\theta)\bm{S}(t)+\bm{N}(t)
\end{equation}
where $\bm{N}(t)\in \mathbb{C}^{L\times M}$ is the additive noise matrix, typically modeled as complex circularly symmetric Gaussian.

In DOA estimation, the covariance matrix of received signals $\bm{\tilde R}_{xx}$ is commonly employed instead of the raw signal matrix $\bm{\tilde X}$ for analysis. This approach offers advantages such as accelerated computation through eigenvalue decomposition (EVD) and dimensionality reduction. $\bm{\tilde R}_{xx}$ is formulated as:
\begin{equation}
\bm{\tilde R}_{xx}=\frac{1}{M}\sum_{t=1}^{M}\bm{\tilde{X}}(t)\bm{\tilde{X}}^H(t)
\label{Rxx}
\end{equation}

% Due to the array manifold limitation of RV-root-MUSIC that only apply to uniform linear arrays(ULA), the array manifold of ULA can be expressed as:
% \begin{equation}
% \bm{A}(\theta)=\begin{bmatrix}
%  \bm{a}(\theta_1) & \bm{a}(\theta_2) & \cdots  & \bm{a}(\theta_k)\\
%  % e^{-j\frac{2\pi}{\lambda {_c}}sin\theta_1 } & e^{-j\frac{2\pi}{\lambda {_c}}sin\theta_2 } & \cdots & e^{-j\frac{2\pi}{\lambda {_c}}sin\theta_K }\\
%  % e^{-j\frac{4\pi}{\lambda {_c}}sin\theta_1 } & e^{-j\frac{4\pi}{\lambda {_c}}sin\theta_2 } & \cdots & e^{-j\frac{4\pi}{\lambda {_c}}sin\theta_K }\\
%  % \vdots  & \vdots & \ddots  & \vdots\\
%  % e^{-j(L-1)\frac{2\pi}{\lambda {_c}}sin\theta_1 }  & e^{-j(L-1)\frac{2\pi}{\lambda {_c}}sin\theta_2 } & \cdots & e^{-j(L-1)\frac{2\pi}{\lambda {_c}}sin\theta_K }
% \end{bmatrix}
% \label{ULA}
% \end{equation}
% where $\bm{a}(\theta _k)$ is the steering vector for the $k$-th DOA $\theta _k$ and is given by:

\subsection{Review about RV-root-MUSIC}
The RV-root-MUSIC algorithm takes the real part of the covariance matrix, where the complex calculation of original root-MUSIC algorithm can be converted into the efficient real-valued one.
Then EVD can be performed to extract the signal and noise subspace:
\begin{equation}
\bm{\tilde R}_{real}=\mathrm{Re}(\bm{\tilde R}_{xx})=\bm{\tilde \mathbb{E}}_s\bm{\tilde \Lambda }_s\bm{\tilde \mathbb{E}}^T_s+\bm{\tilde \mathbb{E}}_n\bm{\tilde \Lambda }_n\bm{\tilde \mathbb{E}}^T_n
\end{equation}
where $span(\bm{\mathbb{E}}_s)$ represents the real-valued eigenvectors spanning the signal subspace, with $\bm{\Lambda }_s$ denoting its associated diagonal eigenvalue matrix. The noise subspace components $span(\bm{\mathbb{E}}_n)$ and $\bm{\Lambda }_n$ are defined analogously using real-valued quantities and $\bm{\tilde R}_{real}=\bm{R}_{real}+\Delta \bm{R}_{real}$, $\bm{\tilde \mathbb{E}}_s=\bm{\mathbb{E}}_s+\Delta \bm{\mathbb{E}}_s$, $\bm{\tilde \Lambda }_s=\bm{\Lambda }_s+\Delta\bm{\Lambda }_s$, $\bm{\tilde \mathbb{E}}_n=\bm{\mathbb{E}}_n+\Delta \bm{\mathbb{E}}_n$, $\bm{\tilde \Lambda }_n=\bm{\Lambda }_n+\Delta\bm{\Lambda }_n=\Delta\bm{\Lambda }_n$. 

Based on the defined noise subspace, the root-finding polynomial for RV-root-MUSIC can be constructed as:
\begin{equation}
f_{RV-root-MUSIC}(z)=\bm{p}^{T}(z^{-1})\bm{\mathbb{E}}_n\bm{\mathbb{E}}^T_n\bm{p}(z)
\label{RVrootMUSIC}
\end{equation}
where $\bm{p}(z)$ is defined as $[1,z,z^2,\cdots ,z^{M-1}]$, $z=e^{jw}$, and $w=\frac{2\pi dsin\theta}{\lambda}$.
Then based on the root-solving results, the DOA estimates can be calculated through:
\begin{equation}
\theta = arcsin[\frac{\lambda}{2\pi d}arg(z)]
\end{equation}

As demonstrated in \cite{FastDOA}, RV-root-MUSIC suffers from the mirror roots, where the relationship $\theta_{mirror} = -\theta_{true}$ is satisfied. Thus the conventional beamforming (CBF) technique is typically employed to filter out the mirror roots as below:
\begin{equation}
\left | \bm{a}^H(\theta)\bm{R}_{xx}\bm{a}(\theta) \right |^2
\end{equation}

\section{Non-asymptotic Performance Analysis of RV-root-MUSIC}
\label{sec:pagestyle}
This paper makes a statistical characteristic analysis of real roots and mirror roots about RV-root-MUSIC under non-asymptotic conditions and also verifies the estimation accuracy with respect to the number of array elements.

The polynomial root-finding function of RV-root-MUSIC can be rewritten as:
\begin{equation}
f=AK(z)G(z)
\label{AKG}
\end{equation}
where $K(z)$ denotes the roots related to the DOA roots:
\begin{equation}
K(z)=(1-r_{k}z^{-1})(1-r_{k}^{*}z)(1-\frac{1}{r_{k}}z^{-1})(1-\frac{1}{r_{k}^{*}}z)
\label{K}
\end{equation}
while $G(z)$ represents other extraneous roots:
\begin{equation}
G(z)=\prod \limits_{j=1,j \ne k}^{\left \lfloor \frac{L-1}{2}\right \rfloor }(1-r_{j}\frac{1}{z})(1-r_{j}^{*}z)(1-\frac{1}{r_{j}}\frac{1}{z})(1-\frac{1}{r_{j}^{*}}z)
\label{G}
\end{equation}

As mentioned, RV-root-MUSIC can provide the mirror roots, therefore we can further divide $K(z)$ into the real root parts $K_{true}(z)=(1-r_{k}z^{-1})(1-r_{k}^{*}z)$ and the mirror root parts $K_{mirror}(z)=(1-\frac{1}{r_{k}}z^{-1})(1-\frac{1}{r_{k}^{*}}z)$:
\begin{equation}
K(z)=K_{true}(z)K_{mirror}(z)=K_{t}(z)K_{m}(z)
\label{Ktm}
\end{equation}

To derive the expression for the DOA estimation bias, we first define a logarithmic function $f(x)=lnx$. Assume that $x=r_k$ is the root corresponding to the $k$-th DOA, and $\Delta r_k$ denotes the error in estimating this root. The first-order Taylor expansion can be performed on $f(x)$ to obtain:
\begin{equation}
f(\tilde r_k)=lnc+j\tilde \omega_k=lnc+j\omega+f'(r_k)\Delta r_k
\label{lnx}
\end{equation}
where $\tilde{r}_k = ce^{j\tilde{\omega}_k} = r_k + \Delta r_k$, $\tilde{\omega}_k = \omega_k + \Delta\omega_k$, and $c$ is the amplitude perturbation caused by additive noise.

Using $f'(r_k)=1/r_k$ and (\ref{lnx}) can be simplified by canceling common terms, then take the imaginary part of both sides, we obtain:
\begin{equation}
\Delta\omega_k = \mathrm{Im}(\frac{\Delta r_k}{r_k})
\label{dwk}
\end{equation}

Note that when $\Delta\theta_k \to 0$, the relationship $\Delta\omega_k =\frac{2\pi d}{\lambda}\cos\theta_k\Delta\theta_k$ is satisfied, we can further derive the bias expression of the final DOA estimation:
\begin{equation}
    \Delta\theta_k = C_k\,\mathrm{Im}(\frac{\Delta r_k}{r_k})
\label{dtk}
\end{equation}
where the scaling factor is $C_k = \frac{\lambda}{2\pi d\cos\theta_k}$. The derivation for mirror DOA follows the same procedure.

\subsection{Deviation of the true DOAs}\label{true}
To derive the perturbation expression, $\mathrm{Im}(\frac{\Delta r_k}{r_k})$ in (\ref{dtk}) is necessary to be computed. The derivation can be decomposed into two steps:

\textbf{step 1}. Take the partial derivative with respect to the root $z$, and substitute $z = r_k$. Considering the perturbation of $r$, replace $r_k$ with $\tilde{r} = r_k + \Delta (r_k)$ and only retain the first-order term to obtain the perturbation expression of $r_k$, as below:
\begin{small}
\begin{equation}
\begin{split}
\left.\frac{\partial f(z,\tilde{r})}{\partial z}\right|_{z=r_k} &= A[K_t(r_k)^{'}[K_m(r_k)G(r_k)]\\
&\quad \left.+K(r_k)[K_m(r_k)G(r_k)]^{'}]\right|_{\tilde{r}=r_k+\Delta r_k}\\ 
                                                                &=A[K_t(r_k)^{'}K_m(r_k)G(r_k)]\\
                                                                &\doteq 2jAr_{k}^{*}Im(\frac{\Delta r_k}{r_{k}})K_m(r_k)G(r_{k})
\end{split}
\label{drt}
\end{equation}
\end{small}

\textbf{step 2}. As in \textbf{step 1}, but now consider the perturbation of $\bm{\mathbb{E}}_n$, replace $\bm{\mathbb{E}}_n$ with $\bm{\tilde \mathbb{E}}_n=\bm{\mathbb{E}}_n+\Delta \bm{\mathbb{E}}_n$ and retain only the first-order terms, we obtain:
\begin{equation}
\begin{split}
\left.\frac{\partial f(z,\bm{\tilde \mathbb{E}}_n)}{\partial z}\right|_{z=r_k} 
&=-r_{k}^{*^{2}}\bm{p}^{(1)}(z^{-1})^{T}\bm{\tilde \mathbb{E}}_n\bm{\tilde \mathbb{E}}_n^{T}\bm{p}(z)\\ &\quad \left.+\bm{p}(z^{-1})^{T}\bm{\tilde \mathbb{E}}_n\bm{\tilde \mathbb{E}}_n^{T}\bm{p}^{(1)}(z)\right|_{z=r_k} \\
\end{split}
\label{dE_n}
\end{equation}
where $p^{(1)}(z)=\frac{\partial p(z)}{\partial z}$. Substitute $\bm{\tilde \mathbb{E}}_n=\bm{\mathbb{E}}_n+\Delta \bm{\mathbb{E}}_n$ and recognize the relation below:
\begin{equation}
\begin{split}
\left\{\begin{array}{l}
\bm{p}^T(\frac{1}{r_k})=\bm{p}^T(r_k^*)=\bm{p}^H(r_k)\\
\bm{p}^{(1)}(\frac{1}{r_k})^T=\bm{p}^{(1)}(r_k^*)^T=\bm{p}^{(1)}(r_k)^H
\end{array}\right.
\end{split}
\label{relation}
\end{equation}

Given that $\bm{\mathbb{E}}^T_n\bm{p}(r_k)=0$ and the relation of (\ref{relation}), (\ref{dE_n}) can be simplified and we can further obtain the perturbation expression of $\bm{\mathbb{E}}_n$ as (\ref{dEt}):
\begin{small}
\begin{equation}
\left.\frac{\partial f(z,\bm{\tilde \mathbb{E}}_n)}{\partial z}\right|_{z=r_k} 
\doteq 2jr_{k}^{*}\mathrm{Im}[r_{k}\bm{p}^{H}(r_{k})\Delta \bm{\mathbb{E}}_n \bm{\mathbb{E}}_n^{T}\bm{p}^{(1)}(r_{k})] \\
\label{dEt}
\end{equation}
\end{small}

Since the perturbation of $r_k$ originates from the perturbation of $\bm{\mathbb{E}}_n$, and both ultimately stem from noise, we can equate (\ref{drt}) and (\ref{dEt}) to obtain:
\begin{equation}
\mathrm{Im}(\frac{\Delta r_k}{r_{k}})=\frac{\mathrm{Im}[r_{k}\bm{p}^{H}(r_{k})\Delta \bm{\mathbb{E}}_n \bm{\mathbb{E}}_n^{T}\bm{p}^{(1)}(r_{k})]}{AK_m(r_k)G(r_{k})}
\label{Imt}
\end{equation}

Substituting to the derived equation (\ref{dtk}) above yields:
\begin{equation}
\Delta \theta _k=\frac{\lambda }{2\pi dcos\theta_k}\frac{\mathrm{Im}[r_{k}\bm{p}^{H}(r_{k})\Delta \bm{\mathbb{E}}_n \bm{\mathbb{E}}_n^{T}\bm{p}^{(1)}(r_{k})]}{AK_m(r_k)G(r_{k})}
\label{totaldtk}
\end{equation}
\subsection{Deviation of the mirror DOAs}\label{mirror}
Since the relationship $r_{\varphi _k}=r_{\theta _k}^*$ is satisfied for the mirror roots $\varphi _k=-\theta _k$, their analysis can be directly obtained by substituting $z=r_{k}^*$ into the derivations from the previous section.

Following the same procedure as in the previous \textbf{step 1}, we can derive the perturbation expression of $r_k^*$:
\begin{small}
\begin{equation}
\begin{split}
\left.\frac{\partial f(z,\tilde{r})}{\partial z}\right |_{z=r_{k}^*} &= A[K_m(r_{k}^*)^{'}[K_t(r_k^*)G(r_{k}^*)]\\
&\quad \left.+K_m(r_{k}^*)[K_t(r_k^*)G(r_{k}^*)]^{'}]\right|_{\tilde{r}=r_{k}^*+\Delta (r_{k}^*)}\\ 
&=\left.A[K_m(r_k^*)^{'}K_t(r_k^*)G(r_k^*)]\right|_{\tilde{r}=r_{k}^*+\Delta (r_{k}^*)}\\
&\doteq 2jAr_{k}Im(\frac{\Delta r_k^*}{r_{k}^{*}})K_t(r_k^*)G(r_{k}^*)
\end{split}
\label{drf}
\end{equation}
\end{small}
and the perturbation expression of $\bm{\mathbb{E}}_n$ as \textbf{step 2}:
\begin{equation}
\begin{split}
\left.\frac{\partial f(z,\bm{\tilde \mathbb{E}}_n)}{\partial z}\right|_{z=r_k^*} 
&=-r_{k}^{2}\bm{p}^{(1)}(z^{-1})^{T}\bm{\tilde \mathbb{E}}_n\bm{\tilde \mathbb{E}}_n^{T}\bm{p}(z)\\ &\quad \left.+\bm{p}(z^{-1})^{T}\bm{\tilde \mathbb{E}}_n\bm{\tilde \mathbb{E}}_n^{T}\bm{p}^{(1)}(z)\right|_{z=r_k^*} \\
\end{split}
\label{dE_n2}
\end{equation}
where $p^{(1)}(z)=\frac{\partial p(z)}{\partial z}$. Substitute $\bm{\tilde \mathbb{E}}_n=\bm{\mathbb{E}}_n+\Delta \bm{\mathbb{E}}_n$ and recognize the relation below:
\begin{equation}
\begin{split}
\left\{\begin{array}{l}
\bm{p}^T(\frac{1}{r_k^*})=\bm{p}^T(r_k)=\bm{p}^H(r_k^*)\\
\bm{p}^{(1)}(\frac{1}{r_k^*})^T=\bm{p}^{(1)}(r_k)^T=\bm{p}^{(1)}(r_k^*)^H
\end{array}\right.
\end{split}
\label{relation2}
\end{equation}

Given that $\bm{\mathbb{E}}^T_n\bm{p}(r_k^{*})=0$ and the relation of (\ref{relation2}), (\ref{dE_n2}) can be simplified and we can further obtain the perturbation expression of $\bm{\mathbb{E}}_n$ as (\ref{dEf}):
\begin{small}
\begin{equation}
\left.\frac{\partial f(z,\bm{\tilde \mathbb{E}}_n)}{\partial z}\right|_{z=r_k^{*}} 
\doteq 2jr_{k}\mathrm{Im}[r_{k}^{*}\bm{p}^{H}(r_{k}^*)\Delta \bm{\mathbb{E}}_n \bm{\mathbb{E}}_n^{T}\bm{p}^{(1)}(r_{k}^{*})] \\
\label{dEf}
\end{equation}
\end{small}

By the equivalence of perturbations, we can get:
\begin{equation}
\mathrm{Im}(\frac{\Delta r_k^*}{r_{k}^{*}})=\frac{\mathrm{Im}[r_{k}^{*}\bm{p}^{H}(r_{k}^*)\Delta \bm{\mathbb{E}}_n \bm{\mathbb{E}}_n^{T}\bm{p}^{(1)}(r_{k}^{*})]}{AK_t(r_k^*)G(r_{k}^{*})}
\label{Imf}
\end{equation}
and further obtain:
\begin{equation}
\Delta \varphi _k=\frac{\lambda}{2\pi dcos\varphi _k}\frac{\mathrm{Im}[r_{k}^{*}\bm{p}^{H}(r_{k}^*)\Delta \bm{\mathbb{E}}_n \bm{\mathbb{E}}_n^{T}\bm{p}^{(1)}(r_{k}^{*})]}{AK_t(r_k^*)G(r_{k}^{*})}
\label{totaldfk}
\end{equation}

It should be noted that $\Delta r_{k}^{*}$ represents the perturbation at $z=r_{k}^*$, while $(\Delta r_{k})^{*}$ denotes the complex conjugate of the perturbation at $z=r_{k}$. These are fundamentally different quantities, as: $\Delta r_{k}^{*} \neq (\Delta r_{k})^{*}$, $(\Delta r_{k}^{*})^{*} \neq \Delta r_{k}$. This reflects how the function's perturbation behavior varies at different evaluation points.
\subsection{Deviation of the Noise subspace}\label{deltaEn}

Both (\ref{totaldtk}) and (\ref{totaldfk}) reveal that biases  $\Delta \theta _k$ and $\Delta \varphi _k$ originate from perturbations in the noise subspace. The noise subspace perturbation is derived as follow:

We define a new signal model $\bm{X}_v=[\bm{X} \hspace{0.5em}\bm{X}^*]$, where the following properties are satisfied  after considering the noise:
\begin{equation}
\begin{split}
\bm{\tilde X}_v \bm{\tilde X}_v^{H}=2\mathrm{Re}(\bm{\tilde X}\bm{\tilde X}^{H})=2M\mathrm{Re}(\bm{\tilde R}_{xx})
\end{split}
\label{XvXvH}
\end{equation}
where $\bm{\tilde X}_v=\bm{ X}_v+\Delta\bm{X}_v$, $\Delta\bm{X}_v=[\Delta\bm{X} \hspace{0.5em}\Delta\bm{X}^*]$. Then Singular Value Decomposition (SVD) of $\bm{\tilde X}_v$ and EVD of $\bm{\tilde R}_{real}$ yield the following results:
\begin{equation}
\begin{split}
\left\{\begin{array}{l}
\bm{\tilde X_v}=\bm{\tilde U}_s \bm{\tilde W}_s \bm{\tilde V}_s^H+\bm{\tilde U}_n \bm{\tilde W}_n \bm{\tilde V}_n^H\\
\bm{\tilde R}_{real}=\bm{\tilde \mathbb{E}}_s\bm{\tilde \Lambda }_s\bm{\tilde \mathbb{E}}^T_s+\bm{\tilde \mathbb{E}}_n\bm{\tilde \Lambda }_n\bm{\tilde \mathbb{E}}^T_n
\end{array}\right.
\end{split}
\label{svdevd}
\end{equation}
where $\bm{\tilde U}_s=\bm{U}_s+\Delta\bm{U}_s$, $\bm{\tilde W}_s=\bm{W}_s+\Delta\bm{W}_s$, $\bm{\tilde V}_s=\bm{V}_s+\Delta\bm{V}_s$, $\bm{\tilde U}_n=\bm{U}_n+\Delta\bm{U}_n$, $\bm{\tilde W}_n=\bm{W}_n+\Delta\bm{W}_n$, $\bm{\tilde V}_n=\bm{V}_n+\Delta\bm{V}_n$.

Substituting (\ref{svdevd}) into (\ref{XvXvH}) can yield:
\begin{equation}
   \bm{\tilde U}_s\bm{\tilde W}_s^2\bm{\tilde U}_s^H+\bm{\tilde U}_n\bm{\tilde W}_n\bm{\tilde W}_n^T \bm{\tilde U}_n^H=2M(\bm{\tilde \mathbb{E}}_s\bm{\tilde \Lambda }_s\bm{\tilde \mathbb{E}}^T_s+\bm{\tilde \mathbb{E}}_n\bm{\tilde \Lambda }_n\bm{\tilde \mathbb{E}}^T_n)
\label{等价特征空间推导}
\end{equation}

Observing (\ref{等价特征空间推导}), we can obtain following relationships:
\begin{small}
\begin{equation}
\begin{split}
    \bm{\tilde U}_s=\bm{\tilde \mathbb{E}}_s ,\bm{\tilde U}_n=\bm{\tilde \mathbb{E}}_n ,\bm{\tilde W}_s^2=2M\bm{\tilde\Lambda}_s ,\bm{\tilde W}_n\bm{\tilde W}_n^T=2M\bm{\tilde\Lambda}_n
\end{split}    
\label{等价特征空间}
\end{equation}
\end{small}

As derived in \cite{LiFu}, for ULA, the perturbation of the noise subspace in the singular value domain of the received signal matrix $\bm{\tilde X}$ can be expressed as:
% \textbf{\textit{Lemma1}}: For ULA, the received signal matrix $\bm{X}$ and its singular value decomposition (SVD) are given by:
% \begin{equation}
% \bm{X}=\bm{U}_s \bm{W}_s \bm{V}_s^H+\bm{U}_o \bm{W}_o \bm{V}_o^H
% \label{Xsvd}
% \end{equation}
% where the noise subspace perturbation can be expressed as:
\begin{equation}
\Delta \bm{U}_o=-\bm{U}_s \bm{W}_s^{-1} \bm{V}_s^{H} \Delta \bm{X}^{H} \bm{U}_o
\label{duo}
\end{equation}

% \textbf{\textit{Proof}}: See Appendix. A.

Based on (\ref{duo}) and the equivalent noise subspace characterization, we derive the noise subspace perturbation for RV-root-MUSIC as:
\begin{equation}
    \Delta \bm{\mathbb{E}}_n=\Delta \bm{U}_n=-\bm{U}_s \bm{W}_s^{-1} \bm{V}_s^{H} \Delta \bm{X}_v^{H} \bm{U}_n
\label{dEn}
\end{equation}
where $\Delta \bm{X}_v=\bm{N}_v=[\Delta \bm{X} \hspace{0.4em} \Delta {\bm{X}}^{*}]=[\bm{N} \hspace{0.5em} \bm{N}^{*}]$.
\subsection{Root Pairs of RV-root-MUSIC in even array elements}\label{num of array}
\begin{figure}[htb]
\setlength{\abovecaptionskip}{0.cm}
\begin{minipage}[b]{0.48\linewidth}
  \centering
  \centerline{\includegraphics[width=4.4cm]{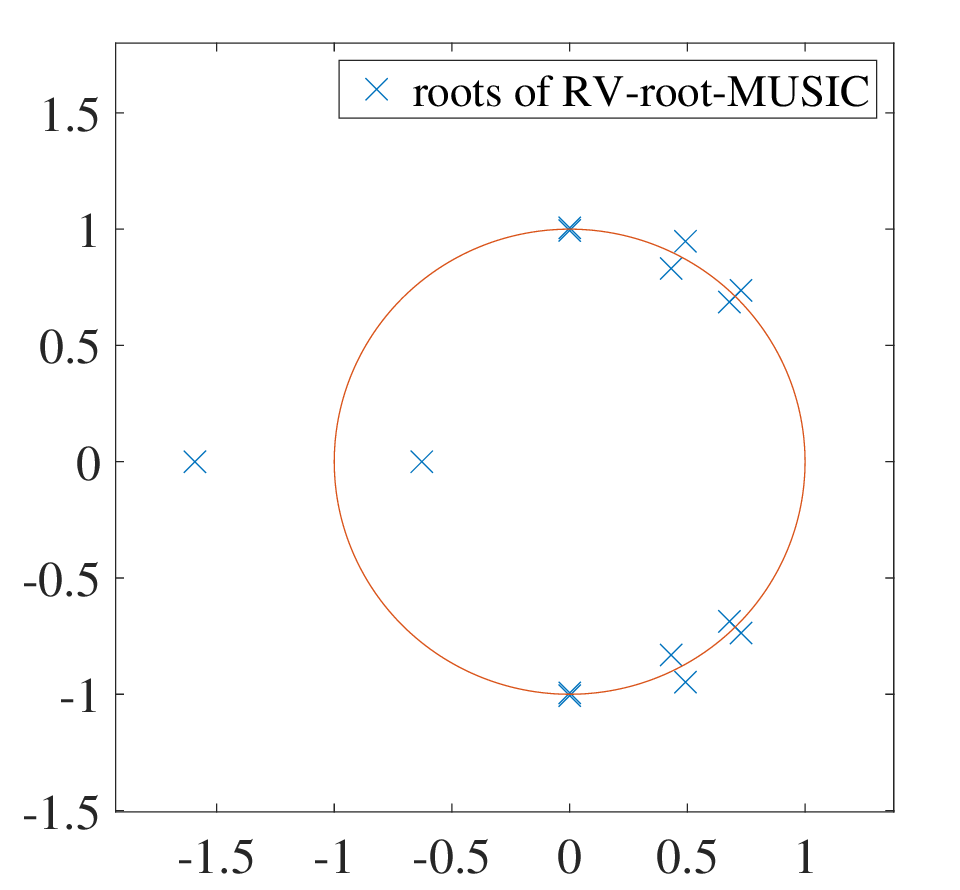}}
%  \vspace{1.5cm}
  \centerline{(a)\quad 8 elements}\medskip
\end{minipage}
\hfill
\begin{minipage}[b]{0.48\linewidth}
  \centering
  \centerline{\includegraphics[width=4.4cm]{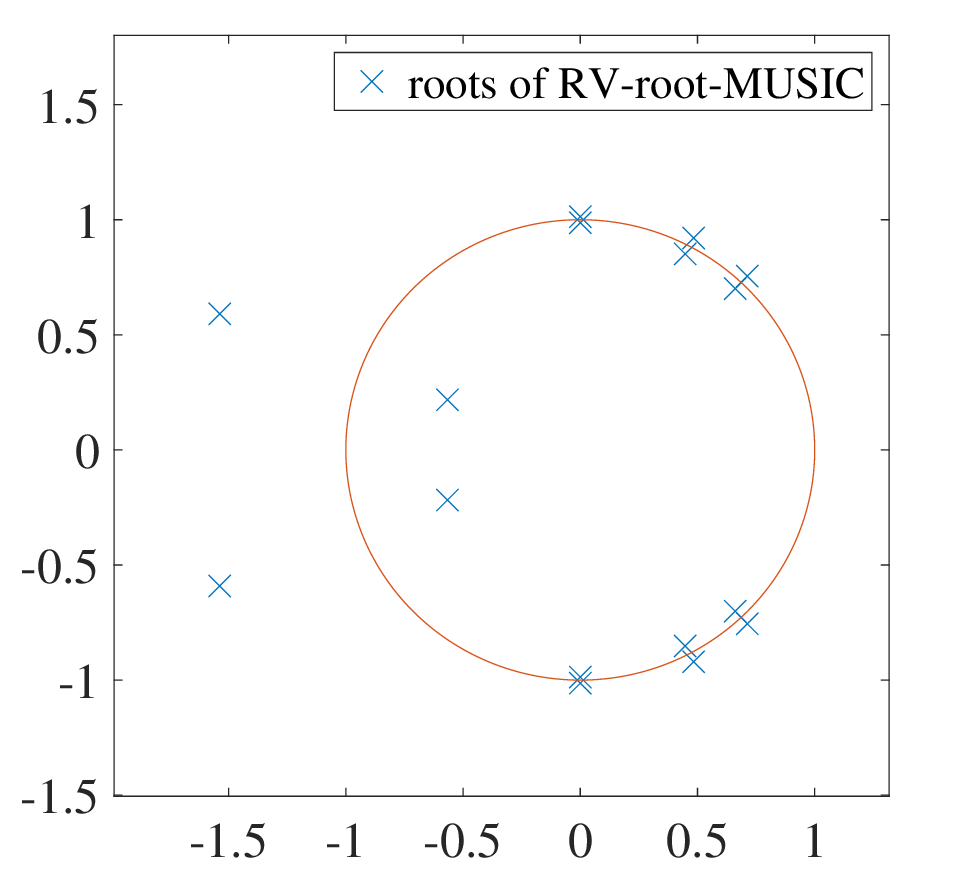}}
%  \vspace{1.5cm}
  \centerline{(b)\quad 9 elements}\medskip
\end{minipage}
\caption{Roots of RV-root-MUSIC in varying array elements}
\label{oden}
\end{figure}

\textbf{\textit{Theorem}}: For ULA with even elements, at least one root pair degenerates onto the real axis, corresponding to spurious DOA at $0^{\circ}$ or $180^{\circ}$ that align with the ULA axis. (see Fig. \ref{oden})
\vspace{-0.3cm}
% The parity of the array element count (odd/even) significantly impacts the root pairing pattern in RV-root-MUSIC. 

% Our analysis reveals(which is shown in Fig. \ref{oden})

% 1. Odd Element Count: 
% Roots exhibit explicit pairing patterns, forming groups of two complex-conjugate pairs.

% 2. Even Element Count:
% One pair of roots degenerates onto the real axis, corresponding to spurious DOA estimates at $0^{\circ}$ or $180^{\circ}$, aligned with the ULA axis. These non-physical solutions are discarded during direction finding.
% \begin{figure}[htbp]
%     \centering
%     \subfigure[\quad 8 elements]{
%         \begin{minipage}{0.225\textwidth}
%             \centering
%             \includegraphics[width=\linewidth, trim=0 0 0 0, clip]{figure/even.eps}
%         \end{minipage}
%     }
%     \subfigure[\quad 9 elements]{
%         \begin{minipage}{0.225\textwidth}
%             \centering
%             \includegraphics[width=\linewidth]{figure/odd.eps}
%         \end{minipage}
%     }
%     \caption{Influence of the parity of the array element}
%     \label{oden}
% \end{figure}
\vspace{-0.3cm}
\noindent
\textbf{\textit{Proof}}:
An odd-degree polynomial can be expressed as:
\begin{equation}
y=a_{n}x^{n}+a_{n-1}x^{n-1}+······+a_{1}x+a_{0}
\label{oddpoly}
\end{equation}
 where $a_{n}\ne 0$ and $n$ is a positive odd integer. Dividing both sides by $x^n$ and taking the limit as $x\to\infty$ yields:
 \begin{equation}
 \lim_{x \to \infty} \frac{y}{x^n} =\lim_{x \to \infty} a_{n}+\frac{a_{n-1}}{x}+\cdots=a_{n}
\label{an}
\end{equation}
Furthermore, since $n$ is odd, we have:
\begin{equation}
\begin{split}
\lim_{x \to +\infty} x^n &=+\infty \quad \lim_{x \to -\infty} x^n =-\infty
\end{split}
\label{2lim}
\end{equation}

By the continuity of $y$ and its behavior at infinity, it can be obtained that there exist points in $\Re$ where $y$ changes sign. The Intermediate Value Theorem guarantees the existence of at least one real root satisfying $y=0$. Extending to the complex domain, for even-numbered arrays, (\ref{RVrootMUSIC}) will transform into an odd-degree polynomial, and there must exist a pair of reciprocal roots on the real axis. \textbf{\textit{Proof}} completed.

Therefore, for the framework proposed in (\ref{G}), we derive a correction factor $R(a)$ for even-numbered arrays as follows:
\begin{equation}
R(a)=a^{2}-2aRe(r_{k})+1=a^{2}-2aRe(r_{k}^*)+1
\label{correction}
\end{equation}
where $a$ is the real root previously discussed. To get more precise perturbation value, just multiply (\ref{totaldtk}) and (\ref{totaldfk}) by $\frac{1}{R(a)}$.

\section{Statistical Properties of True and Mirror DOA Perturbations}
\label{sec:typestyle}
Substituting (\ref{dEn}) into (\ref{totaldtk}) and (\ref{totaldfk}), identical forms can be observed. To simplify the performance analysis model, the general perturbation expression $\Delta _k$ can be expressed as:
\begin{equation}
    \Delta _k=-\frac{\mathrm{Im}(\bm{\beta} _k^{H} \bm{N}_v^{H} \bm{\alpha} _k)}{\gamma _k}    
\label{dk通式}
\end{equation}
where the parameters $\bm{\alpha} _k$, $\bm{\beta} _k$, $\gamma _k$ are specified in Table \ref{abc}.

Given that bias has zero mean and each element of the random variable $\Delta \bm{X}_v$ has variance $\sigma _n^2/2$ for both real and imaginary components, the MSE of bias is as follows:
\begin{small}
\begin{equation}
    E_{\bm{N}_v}(\Delta _k)^2=\frac{E_{\bm{N}_v}[\left | \bm{\beta} _k^{H}\bm{N}_v^{H} \bm{\alpha} _k \right |^2 ]}{2\gamma _k^2}=\frac{\left \| \bm{\alpha} _k \right \|^2 \left \| \bm{\beta} _k \right \|^2 \sigma _n^2}{2\gamma _k^2}
\label{MSE通式}
\end{equation}
\vspace{-0.5cm}
\end{small}
\begin{table}[htbp]\small
\centering
\caption{The parameters in the general expression}
\setlength{\tabcolsep}{6pt} % 调整列间距
\begin{tabular}{>{\raggedright\arraybackslash}p{0.4cm}>{\centering\arraybackslash}p{2.45cm}>{\centering\arraybackslash}p{2.25cm}>{\centering\arraybackslash}p{1.9cm}}
\toprule
    & $\bm{\alpha}_k$ & $\bm{\beta}_k$ & $\gamma_k$ \\
\midrule
$\Delta \theta_k$ & $\bm{\mathbb{E}}_n \bm{\mathbb{E}}_n^{T}\bm{p}^{(1)}(r_{k})C_k r_{k}$ & $\bm{V}_s \bm{W}_s^{-1} \bm{U}_s^{H} \bm{p}(r_{k})$ & $AK_m(r_k)G(r_{k})$ \\
$\Delta \varphi_k$ & $\bm{\mathbb{E}}_n \bm{\mathbb{E}}_n^{T}\bm{p}^{(1)}(r_{k}^{*})C_k r_{k}^{*}$ & $\bm{V}_s \bm{W}_s^{-1} \bm{U}_s^{H} \bm{p}(r_{k}^{*})$ & $AK_t(r_k^*)G(r_{k}^{*})$ \\
\bottomrule
\end{tabular}
\label{abc}
\end{table}
\vspace{-0.4cm}
\section{Simulation Experiments and Results}
\subsection{Experimental Setup and Results}
\label{SM}
To analyze the statistical impact of noise on DOA estimation, 
we conduct Monte Carlo simulations to examine the variation of estimation bias under zero-mean additive white Gaussian noise conditions.

\begin{figure}[htb]
\vspace{-0.1cm}
\setlength{\abovecaptionskip}{0.cm}
\begin{minipage}[b]{0.48\linewidth}
  \centering
  \centerline{\includegraphics[width=4.4cm]{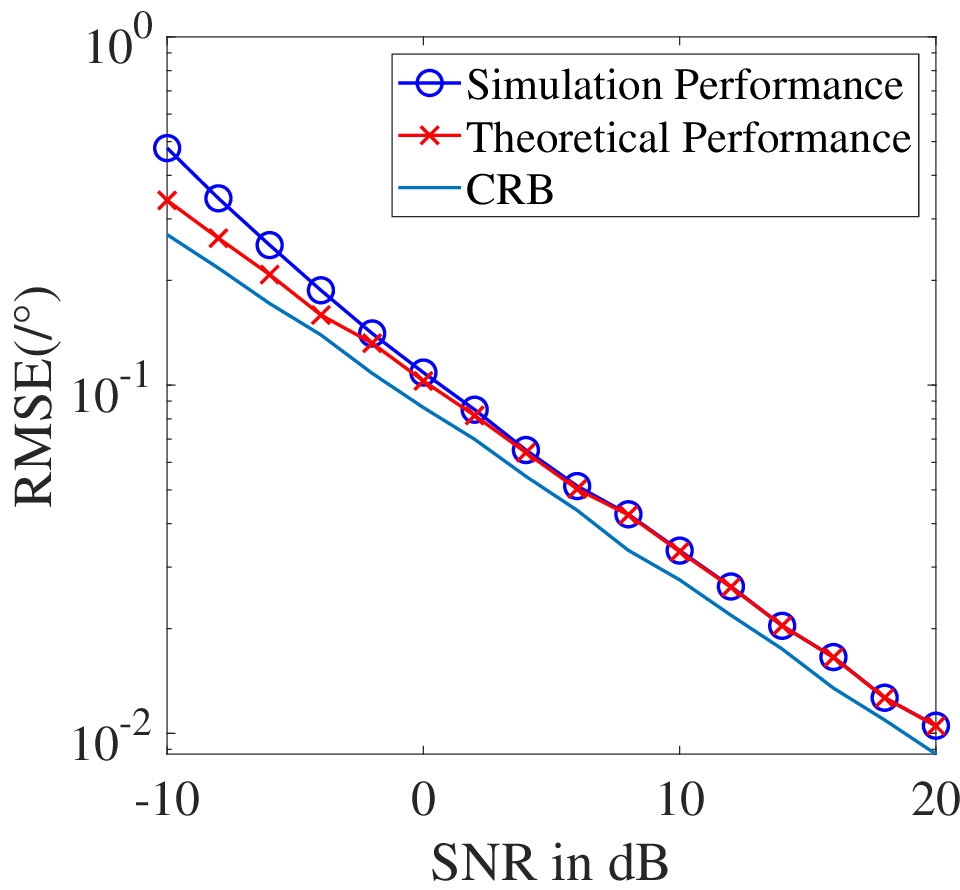}}
%  \vspace{1.5cm}
  \centerline{(a)\quad True DOA}\medskip
\end{minipage}
\hfill
\begin{minipage}[b]{0.48\linewidth}
  \centering
  \centerline{\includegraphics[width=4.4cm]{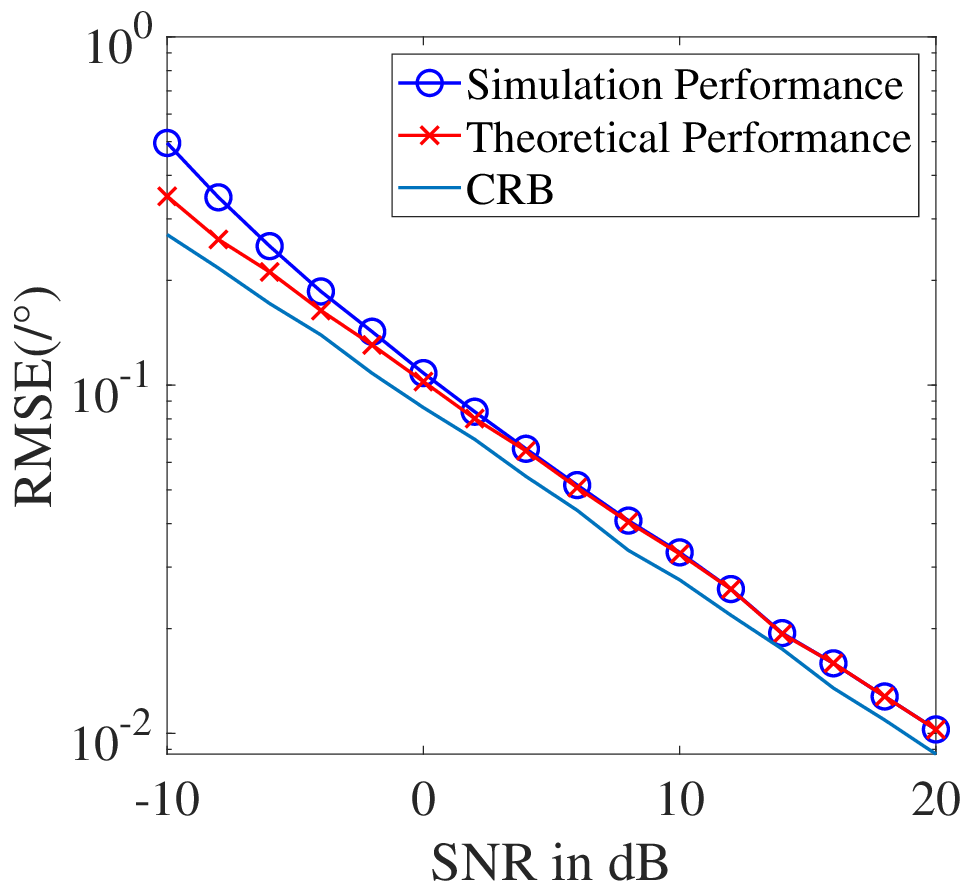}}
%  \vspace{1.5cm}
  \centerline{(b)\quad Mirror DOA}\medskip
\end{minipage}
\caption{The trend of estimation RMSE in SNR variation}
\label{RMSEvSNR}
\end{figure}

\begin{figure}[htb]
\vspace{-0.4cm}
\setlength{\abovecaptionskip}{0.cm}
\begin{minipage}[b]{0.485\linewidth}
  \centering
  \centerline{\includegraphics[width=4.4cm]{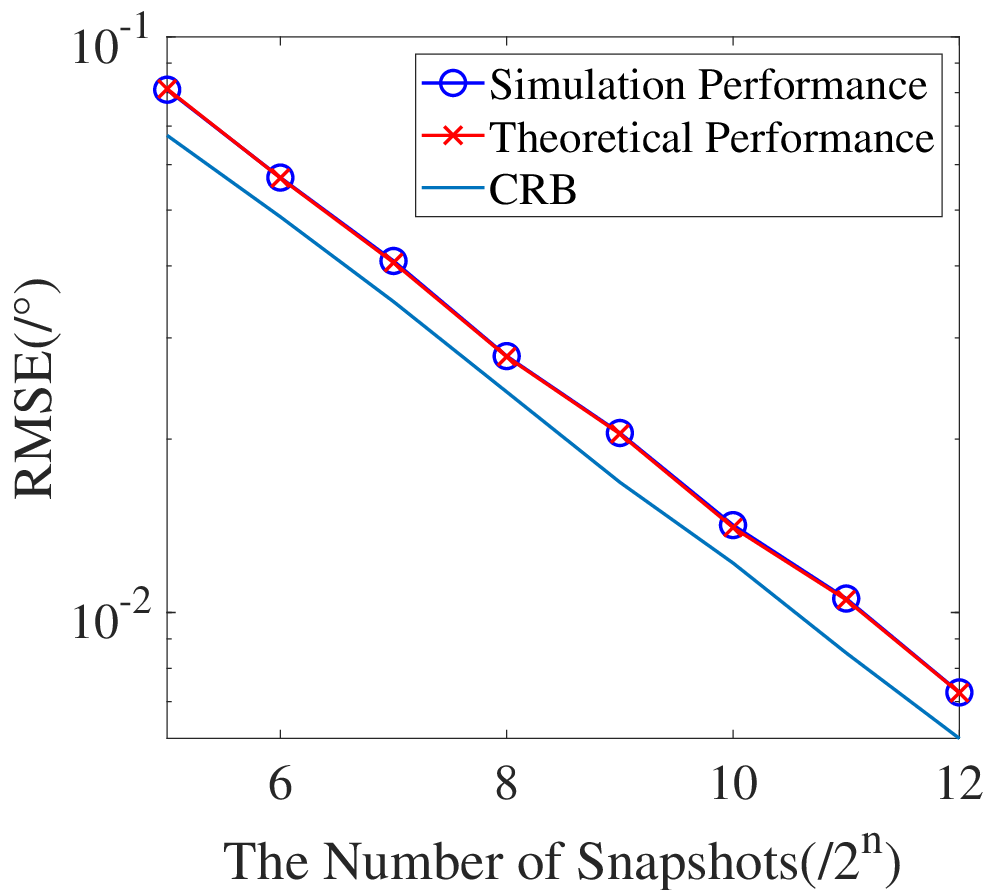}}
%  \vspace{1.5cm}
  \centerline{(a)\quad True DOA}\medskip
\end{minipage}
\hfill
\begin{minipage}[b]{0.485\linewidth}
  \centering
  \centerline{\includegraphics[width=4.4cm]{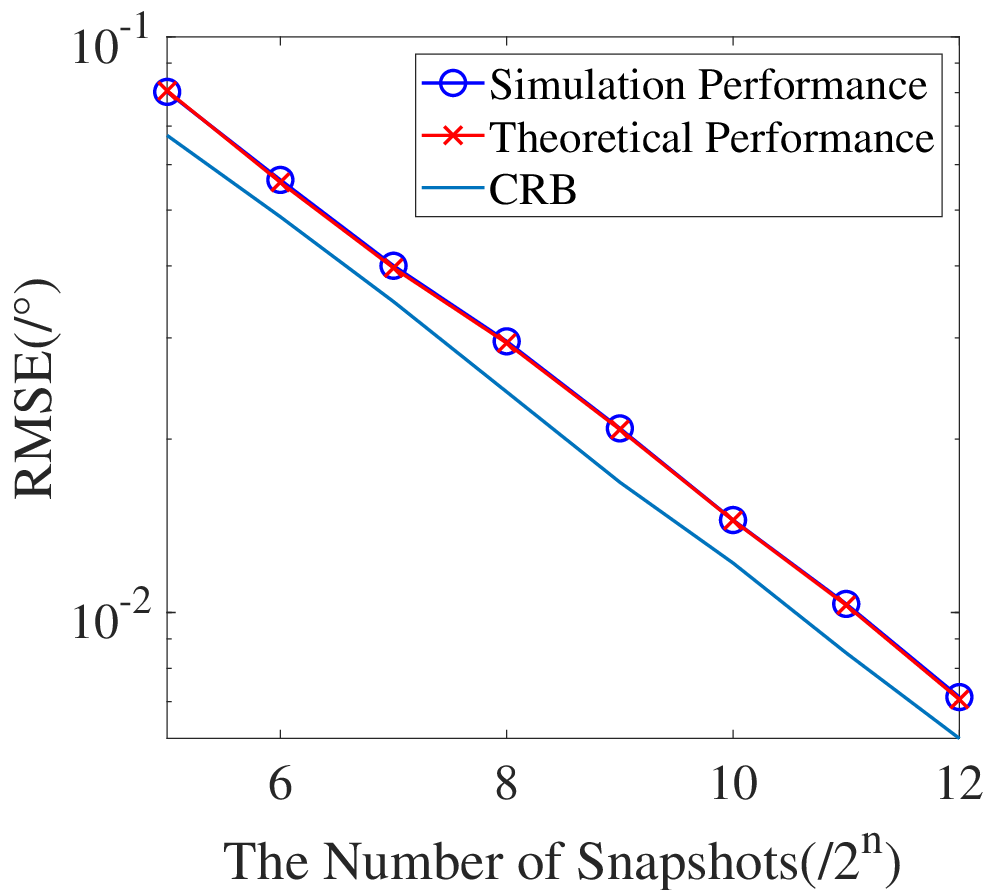}}
%  \vspace{1.5cm}
  \centerline{(b)\quad Mirror DOA}\medskip
\end{minipage}
\caption{The trend of estimation RMSE in Snapshots variation}
\label{RMSEvT}
\end{figure}

% \subsection{Estimation Bias Under Varying SNR Conditions}
% A 9-element ULA with $d=\lambda/2$ is used in the simulation, with two uncorrelated sources at $30^{\circ}$ and $50^{\circ}$, $200$ snapshots, and SNR swept from $0dB$ to $20dB$ in $2dB$ steps. DOA estimation RMSE for the source at $30^{\circ}$ from $1000$ Monte Carlo trials is shown in Fig. \ref{RMSEvSNR}.
A $9$-element ULA with $d=\lambda/2$ is used in the simulation, with two uncorrelated sources at $30^{\circ}$ and $50^{\circ}$. DOA estimation RMSEs for the source at $30^{\circ}$ from $1000$ Monte Carlo trials are investigated under $2$ conditions:\\
\textbf{\textit{condition 1}}: Varying the SNR from $0dB$ to $20dB$ in $2dB$ steps with $200$ snapshots, and the result is shown in Fig. \ref{RMSEvSNR}.\\
\textbf{\textit{condition 2}}: Varying the snapshots $2^n$, where $n$ varies from $5$ to $12$ in $1$ step at $10dB$ SNR, with the result shown in Fig. \ref{RMSEvT}.

From Fig. \ref{RMSEvSNR}, the RMSE decreases as the SNR increases, and the simulation performance aligns with the theoretical performance more and more closely.
\vspace{-0.2cm}
% \begin{figure}[htbp]
%     \centering
%     \subfigure[\quad True DOA]{
%         \begin{minipage}{0.225\textwidth}
%             \centering
%             \includegraphics[width=\linewidth, trim=0 0 0 0, clip]{figure/SNR_true_lar.eps}
%         \end{minipage}
%     }
%     \subfigure[\quad Mirror DOA]{
%         \begin{minipage}{0.225\textwidth}
%             \centering
%             \includegraphics[width=\linewidth]{figure/SNR_mirror_large.eps}
%         \end{minipage}
%     }
%     \caption{The trend of estimation RMSE with SNR variation}
%     \label{RMSEvSNR}
% \end{figure}
% \subsection{Estimation Bias Under Varying Snapshots Number}
% A 9-element ULA with $d=\lambda/2$ is simulated at $10dB$ SNR with two uncorrelated sources at $30^{\circ}$ and $50^{\circ}$. $1000$ Monte Carlo trials of RV-root-MUSIC  are performed with snapshots $2^n$, where $n$ varies from $3$ to $12$ in steps of $1$. The variation of RMSE for the source at $30^{\circ}$ is shown in Fig. \ref{RMSEvT}.

From Fig. \ref{RMSEvT}, the RMSE decreases with increasing snapshot number, though the rate of improvement diminishes progressively. Beyond a certain threshold, further increases in computational complexity and processing time yield marginal performance gains. Consequently, selecting an appropriate snapshot count that balances resolution accuracy and computational efficiency is critical.
\vspace{-0.3cm}
% \begin{figure}[htbp]
%     \centering
%     \subfigure[\quad True DOA]{
%         \begin{minipage}{0.225\textwidth}
%             \centering
%             \includegraphics[width=\linewidth, trim=0 0 0 0, clip]{figure/M_true_large.eps}
%         \end{minipage}
%     }
%     \subfigure[\quad Mirror DOA]{
%         \begin{minipage}{0.225\textwidth}
%             \centering
%             \includegraphics[width=\linewidth]{figure/M_mirror_large.eps}
%         \end{minipage}
%     }
%     \caption{The trend of estimation RMSE with Snapshots variation}
%     \label{RMSEvT}
% \end{figure}

From Fig. \ref{RMSEvSNR} and Fig. \ref{RMSEvT}, it can be seen that the simulation performance closely aligns with the theoretical performance.
\subsection{Experimental Conclusion}
From the above experiments, it can be seen that the trends of the simulation performance and theoretical performance varying with the SNR and the number of snapshots are basically consistent. The theoretical performance provides a tighter theoretical lower bound for RV-root-MUSIC.

% \subsection{Subheadings}
% \label{ssec:subhead}

% Subheadings should appear in lower case (initial word capitalized) in
% boldface.  They should start at the left margin on a separate line.
 
% \subsubsection{Sub-subheadings}
% \label{sssec:subsubhead}

% Sub-subheadings, as in this paragraph, are discouraged. However, if you
% must use them, they should appear in lower case (initial word
% capitalized) and start at the left margin on a separate line, with paragraph
% text beginning on the following line.  They should be in italics.
\section{Conclusion}
\label{sec:print}
% This paper formulated a framework of the non-asymptotic performance analysis of the RV-root-MUSIC algorithm, deeply analyzed the perturbation characteristics of the noise subspace, the true roots and the mirror roots, established a statistical model, and conducted simulation experiments. The final results show that the research in this paper can provide a tighter theoretical lower bound for the performance of the RV-root-MUSIC algorithm. In addition, the study on the parity of the array elements in this paper provides guidance for the performance improvement of the algorithm: that is, choosing an odd number of array elements to ensure the pairing phenomenon of the roots and thereby enhance the robustness of the algorithm.
This paper presented a non-asymptotic performance analysis framework for the RV-root-MUSIC algorithm. Unlike prior works that mainly focus on asymptotic or complex-domain analyses, the proposed framework systematically characterizes the perturbation behavior of the noise subspace, true roots, and mirror roots in the real-valued setting. A statistical model was developed, and generalized perturbation expressions were derived, providing a unified formulation for both true and mirror DOA deviations. Monte Carlo simulations confirmed the theoretical predictions, demonstrating close alignment with the analytical results.

Beyond the theoretical contributions, this work offers several practical insights:
\begin{itemize}
\item The analysis establishes a tighter theoretical performance bound for RV-root-MUSIC in finite-snapshot and low-SNR regimes, which are highly relevant to real-world radar, communications, and sensing systems.
\item The investigation of array element parity revealed that even-element ULAs suffer from root degeneracy on the real axis, leading to spurious DOAs at broadside directions. The proposed correction factor and the recommendation to employ odd-element arrays provide a design guideline for practitioners to enhance algorithm robustness.
\item The derived perturbation expressions enable parameter optimization (e.g., array size, number of snapshots, and SNR trade-offs), offering practical value for system design under constrained computational resources.
\end{itemize}

Overall, this paper advances the theoretical understanding of RV-root-MUSIC under non-asymptotic conditions and bridges the gap between abstract perturbation theory and practical DOA estimation design. Future work will extend the framework to correlated sources, arbitrary array geometries, and adaptive mirror-root suppression techniques, further enhancing its applicability in modern sensing and communication platforms.

\newpage
\section{Funding Acknowledgements}
\label{sec:page}
The work of X. T. Meng is supported by the Shandong Provincial Natural Science Foundation (No.ZR2024QF068). The work of F. Gini and M. Greco has been partially supported by the Italian Ministry of Education and Research (MUR) in the framework of the FoReLab project (Departments of Excellence).

\bibliographystyle{ieeetr}
\bibliography{strings,refs}

\end{document}